\documentclass[12pt]{article}

\usepackage[dvips]{color}                 %%% for colours in the graphs
\usepackage{graphicx}                     %%% for pictures and figures
\usepackage{amssymb}

\usepackage{xspace}                       %%% to get the correct spacing of
\newcommand{\comment}[1]{%
}

\newcommand{\ia}{{\"{\i}}}   %not necessary if \usepackage[T1]{fontenc} is used
\newcommand{\absatz}{\vspace{2ex}\noindent}

\renewcommand{\today}{\ifcase\day\or 1st\or 2nd\or 3rd\or 4th\or 5th\or 6th\or
        7th\or 8th\or 9th\or 10th\or 11th\or 12th\or 13th\or 14th\or 15th\or 
        16th\or 17th\or 18th\or 19th\or 20th\or 21st\or 22nd\or 23rd\or 24th\or
        25th\or 26th\or 27th\or 28th\or 29th\or 30th\or 
        31st\fi~\ifcase\month\or January\or February\or March\or April\or
        May\or June\or July\or August\or September\or October\or November\or
        December\fi \space \number\year}   

\newcommand{\bib}[1]{\bibitem{#1}
    }

\newcommand{\journal}[4]{{#1}\textbf{#2}, #4 (#3)}
\newcommand{\IJMPE}{\textit{Int.\ J.\ Mod.\ Phys.\ }\textbf{E}}

\newcommand{\NPA}{\textit{Nucl.\ Phys.\ }\textbf{A}}
\newcommand{\NPB}{\textit{Nucl.\ Phys.\ }\textbf{B}}
\newcommand{\PLB}{\textit{Phys.\ Lett.\ }\textbf{B}}
\newcommand{\PR}{\textit{Phys.\ Rev.\ }}
\newcommand{\PRep}{\textit{Phys.\ Rept.\ }}

\newcommand{\PRC}{\PR\textbf{C}}
\newcommand{\PRD}{\PR\textbf{D}}

\newcommand{\half}{\frac{1}{2}}
\newcommand{\e}{\mathrm{e}}
\newcommand{\ii}{\mathrm{i}}
\newcommand{\dd}{\mathrm{d}}
\newcommand{\tr}{\mathrm{tr}}
\newcommand{\T}{\mathrm{T}}

\newcommand{\deintdim}[2]{\frac{\dd^{#1}\! #2}{(2\pi)^{#1}}\;}

\newcommand{\lv}{\vec{l}}
\newcommand{\pv}{\vec{\,\!p}\!\:{}}
\newcommand{\qv}{\vec{\,\!q}\!\:{}}

\newcommand{\mpi}{m_\pi}
\newcommand{\fpi}{f_\pi}
\newcommand{\MeV}{\mathrm{MeV}}

\newcommand{\HBCPT}{HB$\chi$PT\xspace}

\newcommand{\calA}{\mathcal{A}} 
\newcommand{\calM}{\mathcal{M}} \newcommand{\calO}{\mathcal{O}}
 \newcommand{\calZ}{\mathcal{Z}}

\newcommand{\mytitle}[1]{
                         \begin{center}
                           \LARGE{\textbf{#1}}
                         \end{center}}
\newcommand{\myauthor}[1]{\textbf{#1}}
\newcommand{\myaddress}[1]{\textit{#1}}
\newcommand{\mypreprint}[1]{\begin{flushright} #1 \end{flushright}}

\begin{document}

\begin{titlepage}
  
  \mypreprint{ \textbf{Final, revised version 13th October 2002}\hfill
    nucl-th/0105048\\
    TUM-T39-01-10}
  
  \vspace*{1cm}

  \mytitle{Pion Deuteron Scattering \\
    Length in Effective Field Theory}
  
  \vspace*{0.5cm}

\begin{center}
  
  \myauthor{Bu\={g}ra Borasoy}\footnote{Email: borasoy@physik.tu-muenchen.de}
  and \myauthor{Harald W.\ Grie\3hammer}\footnote{Email:
    hgrie@physik.tu-muenchen.de}
  
  \vspace*{0.5cm}
  
  \myaddress{
    Institut f{\"u}r Theoretische Physik (T39), Physik-Department,\\
    Technische Universit{\"a}t M{\"u}nchen, D-85747 Garching, Germany}
  
  \vspace*{0.2cm}

\end{center}

\vspace*{0.5cm}

\begin{abstract}
  The $\mathrm{S}$ wave pion deuteron scattering length is presented in an
  effective field theory approach of the two-nucleon system. We include pions
  as dynamical particles which allows us to calculate pion re-scattering
  contributions inside the deuteron. Two-nucleon two-pion contact interactions
  with unknown parameters not determined by chiral symmetry need to be
  introduced in order to renormalise the appearing divergences. By choosing
  their values accordingly we are able to accommodate the available
  experimental data.
\end{abstract}
\vskip 1.0cm
\noindent
\begin{tabular}{rl}
Suggested PACS numbers:& 13.75.G, 14.20.Dh, 21.30.Fe, 25.80.Dj, 27.10.+h\\[1ex]
Suggested Keywords: &\begin{minipage}[t]{11cm}
                    Effective Field Theory, pion nucleon and pion deuteron
                    scattering length, iso-spin even pion nucleon scattering
                    \end{minipage}
\end{tabular}

\vskip 1.0cm

\end{titlepage}

\setcounter{page}{2} \setcounter{footnote}{0} \newpage
  
%%%%%%%%%%%%%%%%%%%%%%%%%%%%%%%%%%%%%%%%%%%%%%%%%%%%%%%%%%%%%%%%%%%%%%%%%%%%%%%
%%%%%%%%%%%%%%%%%%%%%%%%%%%%%%%%%%%%%%%%%%%%%%%%%%%%%%%%%%%%%%%%%%%%%%%%%%%%%%%
%%%%%%%%%%%%%%%%%%%%%%%%%%%%%%%%%%%%%%%%%%%%%%%%%%%%%%%%%%%%%%%%%%%%%%%%%%%%%%%
% Main Body
%

%%%%%%%%%%%%%%% Intro %%%%%%%%%%%%%%%%%%%
\section{Motivation}
\setcounter{equation}{0}
\label{sec:intro}
%%%%%%%%%%%%%%%%%%%%%%%%%%%%

A systematic framework to calculate low-energy scattering processes of hadrons
is provided by Chiral Perturbation Theory ($\chi$PT), the effective field
theory of QCD at low energies.  For example, he determination of the pion
nucleon scattering length has drawn a lot of attention and still remains a
topic of interest.  The chiral corrections to
the Weinberg-Tomozawa current algebra theorem %~\cite{Weinberg:1966kf,Tomozawa}
according to which the iso-scalar (iso-spin even) $\mathrm{S}$ wave pion
nucleon scattering length $a^+:=\half(a_{\pi\,p}+a_{\pi\,n})$ is zero at
leading order (LO) have been calculated by Bernard, Kaiser and
Mei\3ner~\cite{Bernard:1993fp,Bernard:1995pa}.  Their next-to-leading order
(NLO) prediction has recently been refined by determining the unknown
parameters in a comparison of non-zero energy $\pi N$ scattering at the chiral
orders $Q^3$~\cite{Fettes:1998ud} and $Q^4$~\cite{Fettes:2000xg} to Koch's
Karlsruhe~\cite{Koch:1986bn}, Matsinos' EM98~\cite{Matsinos:1997pb} and the
VPI/GW group's SP98~\cite{SAID} partial wave analyses. The obtained range
$a^+_\mathrm{HB\chi PT}=[-0.01\;\dots\;+0.006]\;\mpi^{-1}$ is still compatible
with zero. Recently, Gasser et al.~\cite{Gasser} pointed out that
electro-magnetic iso-spin breaking effects could provide corrections of up to
$7\%$ to the LO result. Because $a^+$ is very small, such effects are stronger
than the na{\ia}ve expectation that they are naturally suppressed by powers of
the fine structure constant $\alpha=1/137$, contributing at most $1\%$.
Another recent \HBCPT analysis of pion nucleon scattering to order $Q^4$
attempting to connect threshold and near-threshold parameters with the low
energy theorems of chiral symmetry found on the basis of~\cite{Koch:1986bn}
that $a^+=-0.008\;\mpi^{-1}$~\cite{Becher:2001hv}.  Direct extractions of
$a^+$ from the same $\pi N$ partial wave analyses yield, on the other hand,
$(+0.0041\pm0.0009)\;\mpi^{-1}$~\cite{Matsinos:1997pb} and
$+0.002\;\mpi^{-1}$~\cite{SAID}.

On the experimental side, the best extraction of $a^+$ comes from measurements
of elastic scattering and single charge exchange in pionic atoms. In
contradistinction to phase shift analyses, no extrapolation to zero energy
scattering is necessary, and electro-magnetic effects are considered small.
Most recently, the ETHZ-Neuch\^{a}tel-PSI~\cite{Schroder:1999uq} collaboration
found $a^+_{\mathrm{exp},\,\pi N}=(-0.0022\pm0.0043)\;\mpi^{-1}$ directly from
the line shift and width change in pionic hydrogen, assuming iso-spin
symmetry.  This is compatible with zero and with the \HBCPT result, but with
smaller error bars. The ongoing experiment R-98.01 at the Paul Scherer
Institute aims to reduce the error by an order of magnitude~\cite{PSIexp}.

The value of the pion deuteron scattering length, which was measured by the
same method as for pionic hydrogen to be $a_{\pi d}=[(-0.0261\pm0.0005)+
\ii(0.0063\pm0.0007)]\;\mpi^{-1}$~\cite{Hauser:1998yd}, has also been used to
constrain $a^+$.  However, since -- as mentioned above -- the deuteron is not
a purely iso-scalar nucleon target, binding and especially pion re-scattering
effects between the two nucleons have to be accounted for. Conventional
potential model approaches which combine deuteron and hydrogen data have been
utilised: Baru and Kudryatsev, e.g., used multiple scattering methods and
quote a value $a^+_\mathrm{phen,\,BK}=
(-0.0015\pm0.0009)\;\mpi^{-1}$~\cite{BaruKu}, however, their errors are
substantially underestimated as shown in~\cite{Ericson:2000md}.  An analysis
by Landau and Thomas~\cite{Thomas:1980xu} gives $a^+_\mathrm{phen,\,LT}=
(+0.0016\pm0.0013)\;\mpi^{-1}$ which recently has been updated by Ericson,
Loiseau and Thomas~\cite{Ericson:2000md} to $a^+_\mathrm{phen,\,ELT}=
(-0.0017\pm0.0002\mathrm{(stat)}\pm0.0008\mathrm{(sys)})\;\mpi^{-1}$.  In that
work it was also mentioned that $a^+$ is also a major source of uncertainty in
the extraction of the coupling constant $g_{\pi NN}$~\cite{Ericson:2000md}.
With the negative value for $a^+$ given in~\cite{Ericson:2000md}, $g^2_{\pi
  NN}/(4\pi)=14.17\pm0.20$ is extracted in contrast to most other approaches
which favour $g^2_{\pi NN}/(4\pi)\approx 13.7\pm0.1$, see Table I
in~\cite{Ericson:2000md}.

Given the unsatisfactory experimental situation for $a^+$, it seems preferable
to take into account binding and pion re-scattering effects between two
nucleons in a more model-independent fashion, in order to constrain $a^+$ from
pion deuteron scattering.  Such a framework is provided by $\chi$PT which
incorporates the chiral symmetry of low-energy QCD.  Employing the so-called
Weinberg counting~\cite{Weinberg} in which the pions are treated
non-perturbatively and iterated to infinite order in ladder exchange diagrams,
Beane et al.~\cite{Beane:1998yg} calculated $a_{\pi d}$. Their result for
$a_{\pi d}$ does not involve any undetermined parameters, however, the
non-perturbative effects responsible for nuclear binding are accounted for
using phenomenological deuteron wavefunctions which introduces an inevitable
model dependence.  It is not clear to what extent the phenomenologically based
deuteron wavefunctions employed in ~\cite{Beane:1998yg} are constrained from
chiral symmetry, and one may therefore
pose the question: \\
Does chiral symmetry predict the pion deuteron scattering length or is
additional phenomenological input needed
in order to fix new unknown parameters? \\
Clearly, this cannot be answered within the Weinberg framework which always
employs deuteron wavefunctions in order to account for the non-perturbative
behaviour of pion exchange.  One must resort to a different approach which is
given by the method of Kaplan, Savage and Wise~\cite{KSW} (ENT(KSW)) where, in
contradistinction to the Weinberg scheme, pion exchange between the two
nucleons is treated perturbatively. It has been illustrated in a number of
papers that ENT(KSW) calculations at next-to-leading order (NLO) are -- for
small momenta -- in good agreement with experimental data. At higher orders,
however, the perturbative expansion is expected to fail in the deuteron
channel~\cite{Fleming:2000ee}.  Since our primary interest is to clarify the
existence and importance of contributions to the pion deuteron scattering
length not fixed by chiral symmetry, we will restrict ourselves to a
next-to-leading calculation in the KSW scheme. As we will see, this issue can
already be investigated at next-to-leading order so that we are not concerned
with problems arising at higher orders.

The article is organised as follows: We start by presenting the Lagrangean
(Sect.~\ref{sec:lagrangean}).  In Sect.~\ref{sec:calculation}, the $\pi d$
scattering amplitude at zero momentum and its scattering length is presented,
followed by a discussion of our findings in Sect.~\ref{sec:results}, including
the comparison to the EFT calculation of Beane et al.~\cite{Beane:1998yg}. The
final Section contains our conclusions. Two Appendices summarise details of
the calculation and renormalisation procedure.

%%%%%%%%%%%%%%% Canonical Stuff %%%%%%%%%%%%%%%%%%%
\section{The Lagrangean}
\setcounter{equation}{0}
\label{sec:lagrangean}
%%%%%%%%%%%%%%%%%%%%%%%%%%%%

We now present the terms of the most general chirally invariant Lagrangean
consisting of contact interactions between non-relativistic nucleons, and
between nucleons and pions, which are relevant for our NLO calculation.

The pertinent terms satisfying the QCD symmetries are in the zero and one
nucleon sector (see e.g.~\cite{Bernard:1995dp})
\begin{eqnarray}
  \label{eq:hbchptlagr}
   \mathcal{L}_{\pi,\,\pi N}&=&\frac{\fpi^2}{8}\;
   \tr[(D_\mu \Sigma^\dagger)( D^\mu \Sigma)]+\;
   \frac{\fpi^2}{4}\;\omega \;\tr [\mathcal{M}_\mathrm{q}
   (\Sigma^\dagger+\Sigma)]
   \;+\nonumber\\
   &&+\;N^\dagger(\ii D_0\;+\;\frac{\vec{D}^2}{2M})N\;+\;
   g_A N^\dagger \vec{A}\cdot\vec{\sigma} N\;+\\
   &&+\;N^\dagger\left[ 2 \omega\; c_1\;
     \tr[\calM_\mathrm{q}(\Sigma+\Sigma^\dagger)]\;+\;4
                  \left(c_2-\frac{g_A^2}{8M}\right) A_0^2\;+\;
                  4 c_3 \;A^\mu A_\mu\right]N\;+\;\dots\;\;,\nonumber
\end{eqnarray}
where $N={p\choose n}$ is the nucleon doublet of two-component spinors,
$M=938.918\;\MeV$ the iso-scalar nucleon mass, and $\sigma$ ($\tau$) the Pauli
matrices acting in spin (iso-spin) space. The field $\xi$ describes the
relativistic pion, for which we choose the sigma gauge for convenience,
\begin{equation}\label{eq:xi}
  \Sigma(x)=\xi^2(x)=\sqrt{1-2\;\frac{\pi^a\,\pi^a}{\fpi^2}}+
  \ii\sqrt{2}\;\frac{\pi^a\,\tau^a}{\fpi}\;\;.
\end{equation}
$D_\mu$ is the chirally covariant derivative $D_\mu=\partial_\mu+V_\mu$, and
the vector and axial currents are
\begin{equation}\label{eq:VandA}
  V_\mu=\frac{1}{2}(\xi\partial_\mu\xi^\dagger+\xi^\dagger\partial_\mu\xi)\;\;,
  \quad
 A_\mu=\frac{\ii}{2}(\xi\partial_\mu\xi^\dagger-\xi^\dagger\partial_\mu\xi)
  \;\;. 
\end{equation}
The pion decay constant is normalised to be $\fpi=130\;\MeV$, $g_A=1.27$,
$\mathcal{M}_\mathrm{q}=\mathrm{diag}(\hat{m},\hat{m})$ is the quark mass
matrix where we work in the iso-spin limit $m_{u}=m_{d}=\hat{m}$, and the
constant $\omega$ is chosen such that $\mpi^2=2 \omega \hat{m}$ at the order
examined here, where we use the iso-scalar value $\mpi=138.039\;\MeV$ for the
pion mass. The coefficients $c_1,\;c_2,\;c_3$ encode high-energy physics
integrated out in \HBCPT and need at present to be determined by experiment.
Since at the scales considered, the momenta of the nucleons are small compared
to their rest mass, the nucleons are treated non-relativistically at leading
order in the velocity expansion, with relativistic corrections systematically
included at higher orders. Thus, the relativistic \HBCPT Lagrangean is reduced
to the form shown above.

The germane terms in the two nucleon Lagrangean are (see
also~\cite{Fleming:2000ee})
\begin{eqnarray}
  \label{eq:kswlagr}
   \mathcal{L}_{NN,\,\pi NN}&=&
   -\;C_0\; (N^\T P^i N)^\dagger \;(N^\T P^i N)\;+\;\frac{C_2}{8}
   \left[(N^\T P^i N)^\dagger\; (N^\T P^i
     (\stackrel{\scriptscriptstyle\rightarrow}{D}-
      \stackrel{\scriptscriptstyle\leftarrow}{D})^2 N)+
   \mathrm{H.c.}\right]\;-\nonumber\\
   &&-\;
    \frac{\omega}{2}\; D_2 \;\tr[\mathcal{M}_\mathrm{q}(\Sigma+\Sigma^\dagger)]
      \;(N^\T P^i N)^\dagger\;(N^T P^i N)\;+\\
   &&+\;\left(E_2\;\tr[A_0^2]\;-\;F_2\;\tr[\vec{A}^2]\right)
      (N^\T P^i N)^\dagger\;(N^T P^i N)
   \;+\;\dots\;\;,\nonumber
\end{eqnarray}
where $P^i$ is the projector onto the $\mathrm{S}$ wave of the
iso-scalar-vector channel,
\begin{equation}\label{eq:proj}
  P^{i,\,b\beta}_{a\alpha}=\frac{1}{\sqrt{8}}\; (\sigma_2\sigma^i)_\alpha^\beta
  \;(\tau_2)_a^b \;\;,
\end{equation}
and the parameters of the Lagrangian $C_i, D_2, E_2$ and $F_2$ are not
constrained by chiral symmetry, but can be extracted from experimental input
or estimated with additional model dependent assumptions.  E.g., $C_0$ can be
related to the binding energy of the deuteron and $C_2$ to the nucleon-nucleon
scattering length~\cite{KSW}.  The remaining parameters $D_2, E_2$ and $F_2$,
on the other hand, have not been determined yet.

%%%%%%%%%%%%%%% Main Part %%%%%%%%%%%%%%%%%%%
\section{Calculation}
\setcounter{equation}{0}
\label{sec:calculation}
%%%%%%%%%%%%%%%%%%%%%%%%%%%%

The pion deuteron scattering length $a_{\pi d}$ follows from the amplitude
$\calA_{\pi d}$ at zero momentum,
\begin{equation}
  \label{eq:pidscattlength}
  a_{\pi d}=\frac{1}{4\pi}\left[1+\frac{\mpi}{2M}\right]^{-1}\;
  \calA_{\pi d}\;\;, 
\end{equation}
which in turn is decomposed into a contribution in which the pion scatters off
only one nucleon (Fig.~\ref{fig:onebody}), and a term with two nucleon
interactions (Fig.~\ref{fig:twobody}),
\begin{equation}
  \label{eq:pidampl}
  \calA_{\pi d}=\calA_{\pi d}^\mathrm{1body}+\calA_{\pi d}^\mathrm{2body}\;.
\end{equation}
The first amplitude starts at LO, $\calO(Q^2)$ after wave function
renormalisation, while the second one is NLO, $\calO(Q^3)$.

%%%%%%%%%%%%%%%%%%%%%%%%%%%%
\subsection{One Body Contributions}
\label{sec:onebody}

\begin{figure}[!htb]
  \centerline{\includegraphics*[width=0.9\textwidth]{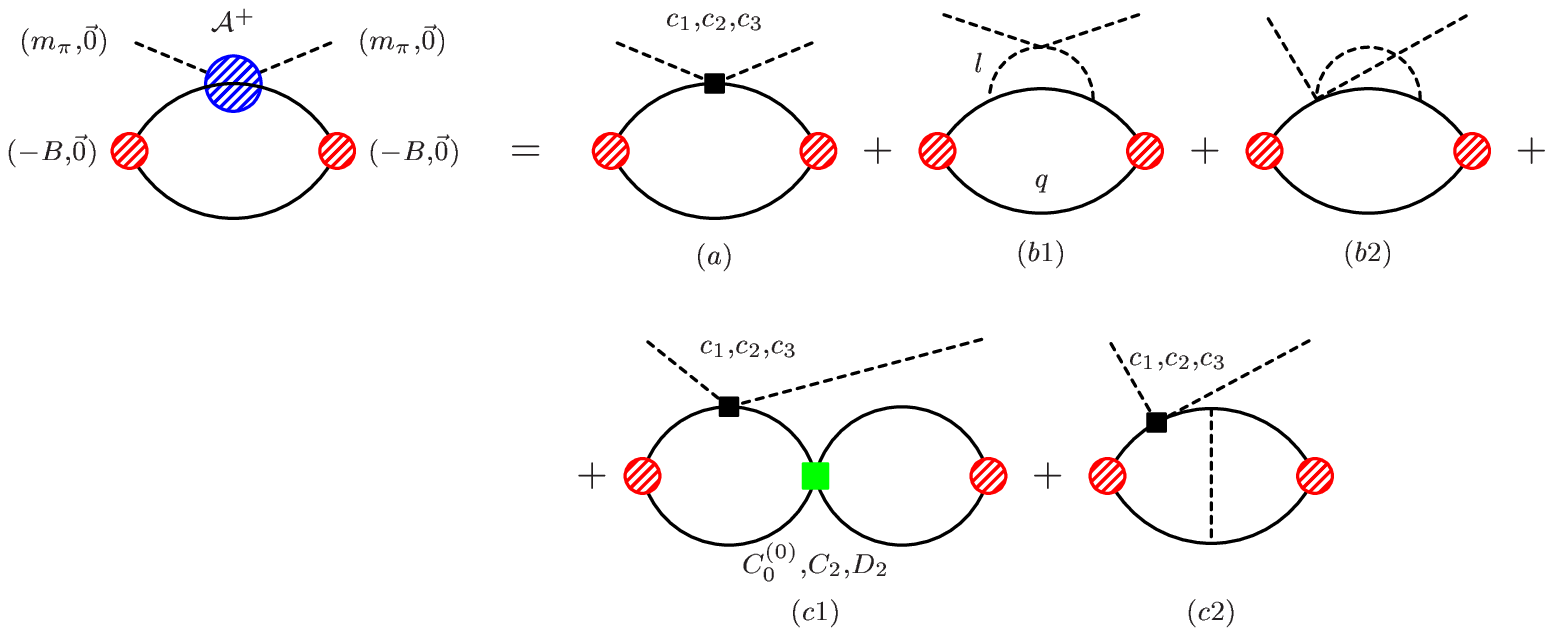}}
    \caption{\textit{The contributions from $\calA^+$: $(a)$: LO; $(b)$: NLO in
        $\calA^+$; $(c)$: NLO deuteron corrections. Shaded circles denote the
        deuteron source. Graphs obtained by permuting vertices or external
        lines are not displayed.}}
    \label{fig:onebody}
\end{figure}

We first discuss briefly how to embed the well known \HBCPT result for the
iso-scalar $\mathrm{S}$ wave pion nucleon scattering amplitude $\calA^+$ into
the deuteron. The one body contributions to $\pi d$ scattering as shown in
Fig.~\ref{fig:onebody} consist of the LO and NLO iso-scalar pion nucleon
amplitude $\calA^+$ at zero momentum (diagrams $(\ref{fig:onebody}a)$ and
$(\ref{fig:onebody}b)$), corrections arising from wave function
renormalisation and NLO deuteron effects (diagrams $(\ref{fig:onebody}c)$).
Since the latter diagrams can be absorbed into a re-definition of the deuteron
source used, it is no surprise that their contribution to $\calA_{\pi d}$ at
zero momentum cancels with the NLO wave function renormalisation of the LO
amplitude.

As the pion can scatter off the proton or neutron inside the deuteron, the one
body amplitude is twice the physical amplitude $\calA^+$ only, independent of
any deuteron observable.  This is formally confirmed in an exemplary
calculation in App.~\ref{sec:AppA}. Relativistic effects do not enter at the
order we calculate. Therefore,
\begin{equation}
  \label{eq:amplitude1body}
  \calA_{\pi d}^\mathrm{1body}=\frac{4\mpi^2}{\fpi^2}\;
  \left[\frac{\Delta_\mathrm{bare}(\mu)}{2}-\frac{g_A^2}{8M}
         +\frac{g_A^2 (3\mpi-2\mu)}{64 \pi \fpi^2}\right]=2\calA^+\;\;
\end{equation}
with the bare quantity $\Delta_\mathrm{bare}(\mu):=
2\left(c_{2,\,\mathrm{bare}}(\mu)+c_{3,\,\mathrm{bare}}(\mu)
  -2c_{1,\,\mathrm{bare}}(\mu)\right)$.  Notice that both graphs $(b1)$ and
$(b2)$ in Fig.~\ref{fig:onebody} depend on the renormalisation scale $\mu$, as
the nested integral is linearly divergent. In the $\textrm{MS}$ or
$\overline{\textrm{MS}}$ scheme usually chosen in \HBCPT, the divergence is
discarded ($\Delta_\mathrm{bare}(\mu)=\Delta^{\overline{\mathrm{MS}}}$), but
it is manifest in the PDS scheme which for consistency has of course to be
used also in the one nucleon part of the calculation. Therefore, the \HBCPT
parameters $c_1,\;c_2,\;c_3$ start to depend on the renormalisation procedure
at NLO and are not observables.  It is however obvious that the regulator
dependence is absorbed into the NLO, $\calO(Q)$ part of the combination of
coefficients $\Delta(\mu)=\Delta^{(0)}+\Delta^{(1)}(\mu)$, in agreement with
the power counting. The LO part, $\Delta^{(0)}$, is $\mu$ independent. The
value for $\Delta$ in the $\overline{\textrm{MS}}$ scheme is hence easily
translated into the PDS scheme as $\Delta_\mathrm{bare}(\mu)
=\Delta^{\overline{\mathrm{MS}}}+\frac{g_A^2\mu}{16\pi\fpi^2}$, making the
full, physical one body amplitude (\ref{eq:amplitude1body}) explicitly $\mu$
independent.

Recently, updated values for the parameters $c_i$ have been determined in an
$\calO(Q^3)$ and $\calO(Q^4)$ \HBCPT fit to three finite energy pion nucleon
scattering analyses~\cite{Fettes:1998ud,Fettes:2000xg}. As already noted in
Refs.~\cite{Bernard:1993fp,Bernard:1995dp}, the iso-scalar $\mathrm{S}$ wave
scattering length
\begin{equation}
  \label{eq:aplus}
  a^+:=
    \frac{1}{4\pi}\;\left[1+\frac{\mpi}{M}\right]^{-1}\;\calA^+
\end{equation}  
can however not be determined precisely because of a numerical cancellation
which may signal physics at small scales not yet understood. The \HBCPT
analysis predicts the range~\cite{Fettes:1998ud,Fettes:2000xg}
\begin{equation}
  \label{eq:aplusvalue}
  a^+_{\mathrm{HB\chi PT}}
  =[-0.01\dots+0.006]\;\mpi^{-1}=(-0.002\pm0.008)\;\mpi^{-1}\;\;,
\end{equation}
compatible with the various experimental and phenomenological extractions
discussed in the Introduction. The comparatively large range comes from the
use of different partial wave analyses for the $\pi N$ amplitudes; the
theoretical uncertainty from \HBCPT is considerably smaller.

%%%%%%%%%%%%%%%%%%%%%%%%%%%%
\subsection{Two Body Contributions}
\label{sec:twobody}

\begin{figure}[!htb]
  \centerline{\includegraphics*[width=0.8\textwidth]{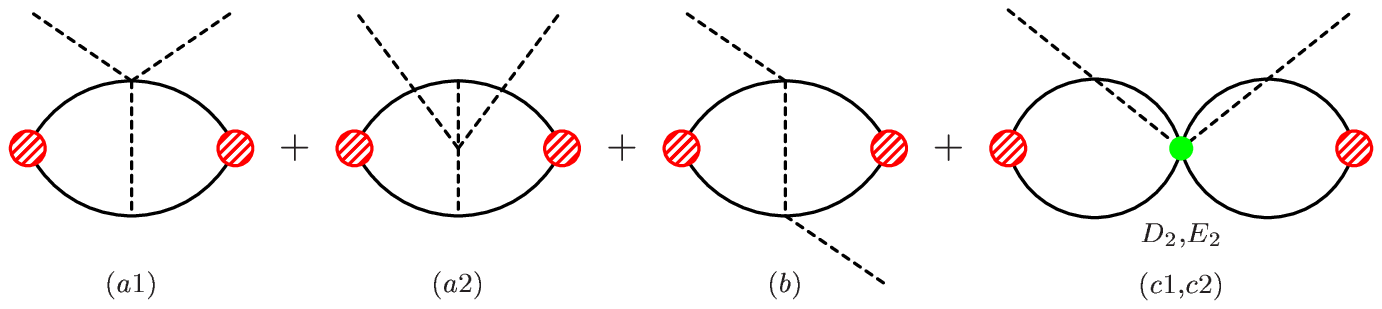}}
    \caption{\textit{The two body contributions from pion re-scattering
        $(a,b)$ and two body contact interactions $(c)$ at NLO. Graphs
        obtained by permuting vertices or external lines are not displayed.}}
    \label{fig:twobody}
\end{figure}

Processes involving deuteron correlations, Fig.~\ref{fig:twobody}, enter at
NLO, $\calO(Q^3)$. The sum of the two diagrams Fig.~\ref{fig:twobody} $(a1)$
and $(a2)$ is independent of the parametrisation of the pion field. The two
pion one nucleon vertices in $(b)$ stem from the chirally covariant form of
the time derivative in the kinetic energy term for the nucleon in
(\ref{eq:hbchptlagr}).

The diagrams with a pion in the intermediate state, Figs.~\ref{fig:twobody}
$(a)$ and $(b)$, contain logarithmic divergences. In order to see how they
manifest themselves in a more traditional setting, consider the graphs when a
deuteron wave function is used instead of the point-like source with the
correct quantum numbers which is the starting point of the standard field
theoretical treatment. In this case, the contributions are proportional to the
expectation values of two operators~\cite{Beane:1998yg,Weinberg:1992yk}
\begin{eqnarray}
  \label{eq:tradpioncorr}
  &&\calA^{\mathrm{2body}\,(a)}_{\pi d,\;\mathrm{trad}}\propto
  \langle \frac{(\qv\cdot\vec{\sigma}_1)(\qv\cdot\vec{\sigma}_2)}{
    (\qv^2+\mpi^2)^2}\rangle\;\;,
    \nonumber\\
  &&\calA^{\mathrm{2body}\,(b)}_{\pi d,\;\mathrm{trad}}\propto
  \langle \frac{1}{\qv^2}\rangle\;\;
\end{eqnarray}
between deuteron wave functions, where $\qv$ is the momentum transfer between
the nucleons. Thus, the graphs seem to probe the deuteron only at momenta of
the order of $q\sim\mpi$ in Fig.~\ref{fig:twobody} $(a)$ and $q\sim 0$ in
Fig.~\ref{fig:twobody} $(b)$. The Fourier transformation into position space
reveals however that the expectation values of the operators $\frac{1}{r}$ and
$\frac{\e^{-\mpi r}}{r}$ are probed, testing the deuteron wave function at
arbitrarily short distance. Usually, the UV part of the deuteron wave function
is parameterised by adding to the long range pionic potentials strong
phenomenological terms mimicking short range repulsion. The deuteron wave
function is thus suppressed at large momentum transfer at the price of adding
some dependence on the shape and parameters of the short distance potential.
Thus, the result for these amplitudes depends also on short distance physics
subsumed into an arbitrary, unphysical cut-off parameter.

It is therefore no surprise that a logarithmic divergence appears in the pion
exchange diagrams.  The necessary integrals are tabulated
in~\cite{Savage:2001cm}, so that the bare amplitudes are
\begin{eqnarray}
  \label{eq:barepioncorr}
  &&\calA^{\mathrm{2body}\,(a)}_{\pi d,\;\mathrm{bare}}=
    \frac{g_A^2\mpi^2\gamma}{12\pi\fpi^4}
    \left(\Gamma-\;\frac{\mpi}{\mpi+2\gamma}\;-
      2\ln\left[\frac{\mpi+2\gamma}{\mu}\right]\right)
    \nonumber\\
  &&\calA^{\mathrm{2body}\,(b)}_{\pi d,\;\mathrm{bare}}=-\;
  \frac{\mpi^2\gamma}{\pi\fpi^4}
    \left(\Gamma-2\ln\left[\frac{2\gamma}{\mu}\right]\right)
\end{eqnarray}
with a divergence $\Gamma:=\Gamma[4-d]+\ln[\pi]+1$ in $d\to 4$ space-time
dimensions~\footnote{Following~\cite{Savage:2001cm}, we prefer to work with a
  definition of $\Gamma$ which is independent of the scale $\mu$, in
  contradistinction to e.g.~\cite{Bernard:1995dp}.}. Therefore, the bare two
pion two nucleon contact interactions Fig.~\ref{fig:twobody} $(c1)$ and $(c2)$
entering at the same order are also necessary in order to consistently remove
all regulator dependence in the total, physical two body amplitude.
\begin{eqnarray}
  \label{eq:bare2bodycorra}
  &&\calA^{\mathrm{2body}\,(c)}_{\pi d,\;\mathrm{bare}}=
  \frac{\mpi^2\gamma}{\pi\fpi^2}\;
  \left[D_{2,\;\mathrm{bare}}(\mu)+E_{2,\;\mathrm{bare}}(\mu)\right]
  (\gamma-\mu)^2 
\end{eqnarray}
They serve the same purpose as suppressing the deuteron wave function at short
distance by a cut-off.  As must be expected from the power counting, the
combinations $D_{2,\;\mathrm{bare}}(\mu)\;(\gamma-\mu)^2$ and
$E_{2,\;\mathrm{bare}}(\mu)\;(\gamma-\mu)^2$ scale as $Q^0$. The strengths of
both two pion two nucleon contact terms cannot be predicted at the present
time. Albeit the parameter $D_2$ is encountered as the chiral symmetry
breaking contribution to $NN$ scattering at NLO, in this process its
renormalised value cannot be dis-entangled from the parameter $C_0^{(0)}$
which respects chiral symmetry~\cite{Fleming:2000ee}. The strength $E_2$ is
not fixed, either.  Unfortunately, we are also not able to extract their
values from additional experimental input so that we cannot predict the pion
deuteron scattering length.

Nonetheless, EFT allows to estimate the natural size of the combination
$D_2+E_2$ after specifying a prescription to handle the divergences in three
and four dimensions. Only after renormalisation is one justified to compare
the sizes of the various contributions. In the PDS scheme, the theory is
usually renormalised (and the power counting made manifest even before
renormalisation) by choosing the renormalisation scale to be
natural~\cite{KSW}:
\begin{equation}
  \label{eq:naiverenormalisation}
  \mu\to\mpi\;\;,\;\;\Gamma\to2
\end{equation}
With the latter choice, the divergences in $4$ dimensions are removed by
demanding that the pionic contribution to zero momentum scattering between two
nucleons disappears, see (\ref{eq:NNonepion}) in App.~\ref{sec:AppB}.  Power
counting and dimensional analysis dictate then that
\begin{equation}
  \label{eq:D2E2natural}
  D_2+E_2=\frac{4\pi}{M\;\Lambda_{NN}}\;\frac{\calZ}{(\mu-\gamma)^2},
\end{equation}
where the magnitude of the dimensionless parameter is of order unity,
$|\calZ|\approx1$, if the naturalness assumption holds. Its sign remains
undetermined. The scale $\Lambda_{NN} \approx 300$ MeV enters in the chiral
expansion for $NN$ scattering \cite{KSW} and subsumes al short distance
physics, i.e.~all effects of particles not contained as explicit degrees of
freedom in the Lagrangean, like the $\Delta$, $\rho$ meson exchange etc.
%%%**********************************************************

However, it must be stressed that in the approach taken here, a decomposition
of the physically observable $\pi d$ scattering length into parts related to
the pion re-scattering diagrams of Fig.~\ref{fig:twobody} $(a)$ and $(b)$
separately is strictly speaking meaningless: None of these diagrams is
renormalisation group invariant (i.e.~cut-off independent) on its own, and
only combinations of these diagrams with the two pion two nucleon contact
diagrams of Fig.~\ref{fig:twobody} $(c)$ form observables free of divergences,
i.e.~independent of $\mu$ and $\Gamma$.  After renormalisation performed in
App.~\ref{sec:AppB}, the scattering amplitude reads
\begin{equation}
  \label{eq:twobodyRGinv}
  \calA_{\pi d}^\mathrm{2body}=\frac{\mpi^2 \gamma}{\pi\fpi^4}\;
  \left[2\ln\left[\frac{2\gamma}{\Lambda^*}\right]-
    \frac{g_A^2}{6}\left(\half\;\frac{\mpi}{\mpi+2\gamma}+
      \ln\left[\frac{(\mpi+2\gamma)}{\Lambda^*}\right]
    \right)\right]\;\;,
\end{equation}
and contains only one un-determined, physical parameter $\Lambda^*$. This
dimension-ful number parametrises the renormalisation group invariant strength
of the contact interactions $D_2$ and $E_2$ between nucleons and subsumes
effects from the deuteron wave function at short distances. It needs to be
determined from experiment or from a microscopic calculation of $NN$
scattering in QCD.  $\Lambda^*$ is expected to be of the order of the natural
low energy scale ($\mpi$ or $\Lambda_\mathrm{QCD}$) since all dependence on
higher scales integrated out has disappeared with renormalisation.

As mentioned above, relativistic effects do not enter at the order we are
working. Nevertheless, it may be worthwhile to comment on their contributions.
Relativistic corrections to the energy-momentum relation are accounted for by
inserting perturbatively higher dimension operators, the lowest one being
$\pv^4/(8M^3)$. They start at N2LO and are suppressed by factors of the
nucleon mass, and not by $\Lambda_{NN} \approx 300$ MeV.  Such effects are
therefore small compared to other corrections which enter formally at the same
order N2LO. Our calculation only includes NLO effects. For our purposes, they
are therefore negligible, and may be regarded to lie within the given error
bars.
%%%***************************

%%%%%%%%%%%%%%% Results %%%%%%%%%%%%%%%%%%%
\section{Results and Discussion}
\setcounter{equation}{0}
\label{sec:results}
%%%%%%%%%%%%%%%%%%%%%%%%%%%%

With the data at hand, we cannot predict either the iso-scalar or the pion
deuteron scattering length in a unique way from our calculation due to the
unknown physical scale $\Lambda^*$ stemming from the combination $D_2+E_2$ of
counter terms. These unknown strengths of the two-pion two-nucleon couplings
enter already at NLO, i.e.~at the same order as pion re-scattering.  In order
to determine their sizes, we use as inputs the experimental value for the pion
deuteron scattering length and a value for $a^+$. Clearly, requiring
self-consistency forbids to consider phenomenological extractions of $a^+$
into which $a_{\pi d}$ entered. We use as choice either the value from the
pionic hydrogen experiment, $a^+_{\mathrm{exp},\,\pi
  N}=(-0.0022\pm0.0043)\;\mpi^{-1}$~\cite{Schroder:1999uq}, or the \HBCPT
prediction, Eq.~(\ref{eq:aplusvalue})~\cite{Fettes:1998ud,Fettes:2000xg}. Both
are compatible with zero, and the \HBCPT result has an error bar accommodating
also the phenomenological partial wave analysis
extractions~\cite{Koch:1986bn,Matsinos:1997pb,SAID} cited in the Introduction.
We find
\begin{eqnarray}
  \label{eq:Lambdavalue}
  \Lambda^*= (262^{+185}_{-109})\;\MeV&& \mbox{ when fitted to }
       a^+_{\mathrm{exp},\,\pi N}\mbox{~\cite{Schroder:1999uq}}
  \nonumber\\
  \Lambda^*=(269^{+458}_{-170})\;\MeV&&\mbox{ when fitted to }
  a^+_\mathrm{HB\chi
    PT}\mbox{~\cite{Fettes:1998ud}}\;\;,
\end{eqnarray}
which contains a large error bar from the uncertainty in $a^+$. The value of
$\Lambda^*$ increases as $a^+$ increases. The Logarithms in
(\ref{eq:twobodyRGinv}) are indeed of order $1$.

\begin{figure}[!htb]
  \begin{center}
    \includegraphics*[width=0.6\textwidth]{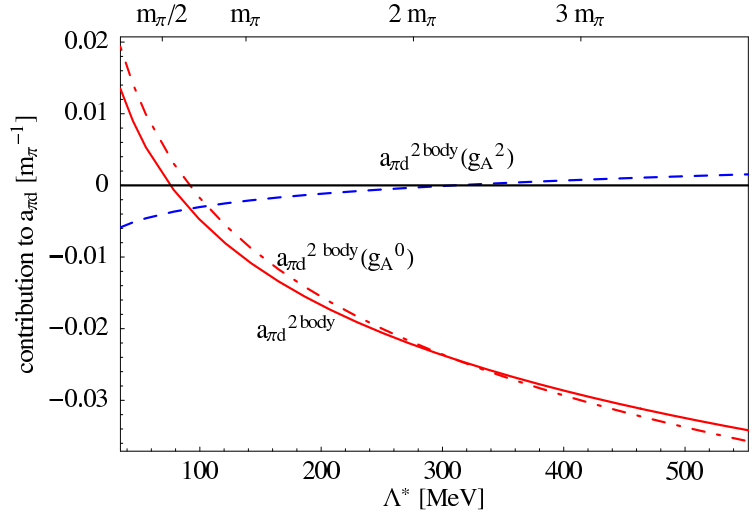}
\caption{\textit{Sizes of the renormalisation group invariant contributions to
    $a_{\pi d}$, (\ref{eq:twobodyRGinv}), depending on the physical parameter
    $\Lambda^*$. Dashed line: contribution binomial in $g_A$; dash-dotted
    line: contribution independent of $g_A$; solid line: total.}}
\label{fig:twobodycontribs}
  \end{center} 
  \vspace*{-4ex}
\end{figure}

We now consider the sizes of two renormalisation group invariant subsets
constructed out of the two body scattering result (\ref{eq:twobodyRGinv}): The
first one is the term independent of $g_A$. The second one is quadratic in
$g_A$, and is easily estimated to be suppressed by a factor $1/12$.  We
reproduce the well known result that the dominant pion re-scattering
contribution stems from physics unchanged by taking the chiral limit, and
$\Lambda^*\approx\Lambda_{E_2}^*$.  Albeit the definition of $\Lambda^*$
depends also on $g_A$ (\ref{eq:Lambda}), this polynomial separation with
respect to $g_A$ is at present a good analogue to the values quoted in the
Weinberg approach for the pion re-scattering contributions,
Fig.~\ref{fig:twobody} (a) and (b). Although much simpler, our results for
these renormalisation group invariant combinations are close to the findings
of the Weinberg approach, see Table~\ref{tab:comparetoWeinberg}.  This should
not be too much a surprise since it was shown above that the pion
re-scattering diagrams test after renormalisation only the tail of the
deuteron wave function, at momenta not larger than $\mpi$.
Figure~\ref{fig:twobodycontribs} depicts the dependence of the two body
contributions to the $\pi d$ scattering length on $\Lambda^*$.  We finally
summarise all one and two body contributions to $a_{\pi d}$ in
Table~\ref{tab:twobody} for the extraction using $a^+_{\mathrm{exp},\,\pi N}$.

\begin{table}[!htb]
    \begin{center}
      \begin{tabular}{|l|l|l|}
        \hline
        approach \rule{0ex}{15\unitlength}
                 & $a_{\pi d}^{\mathrm{2body}}(g_A^2)$ $[\mpi^{-1}]$
                 & $a_{\pi d}^{\mathrm{2body}}(g_A^0)$ $[\mpi^{-1}]$\\[1ex]
        \hline
        KSW , RG invariant \rule{0ex}{15\unitlength}&
                                              $-0.0004%4
                                                        $ &
                                                      $-0.0210%095
                                                                  $\\[1ex]
        \hline
        Weinberg \rule{0ex}{15\unitlength}&
                                              $-0.0007\pm0.0002$ &
                                                      $-0.0196\pm0.0005$\\[1ex]
%        Weinberg P.C.~with Bonn B \cite{Machleidt:1989tm}
%               \rule{0ex}{15\unitlength}&
%                                              $-0.000575$ & $-0.02021$\\[1ex]
%        Weinberg P.C.~with ANL-V18 \cite{Wiringa:1995wb}&
%                                              $-0.000792$ & $-0.01960$\\[1ex]
%        Weinberg P.C.~with Reid-SC \cite{Reid:1968sq} &
%                                              $-0.000850$ & $-0.01942$\\[1ex]
%        Weinberg P.C.~with SSC \cite{SSC}&
%                                              $-0.000699$ & $-0.01920$\\[1ex]
        \hline
      \end{tabular}
      \caption{\textit{Comparison of two body corrections to $a_{\pi d}$ in
          the KSW power counting with $\Lambda^*$ determined from
          $a^+_{\mathrm{exp},\,\pi N}$~\cite{Schroder:1999uq}, and in the
          hybrid approach based on Weinberg's power counting taken
          from~\cite{Beane:1998yg}. Error bars from the uncertainty in
          $a^+_{\mathrm{exp},\,\pi N}$ omitted. The error bars in the Weinberg
          approach stem from using different model wave functions for the
          deuteron.}}
      \label{tab:comparetoWeinberg}
    \end{center}
  \vspace*{-4ex}
\end{table}

\begin{table}[!htb]
\vspace*{4ex}  
    \begin{center}
      \begin{tabular}{|l|l|}
        \hline
        diagram \rule{0ex}{15\unitlength} &
           \rule{4em}{0em}contribution to
           $a_{\pi d}$ [$\mpi^{-1}$]\rule{4em}{0em}\\[1ex]
        \hline  
        1 body from $a^+_{\mathrm{exp},\,\pi N}$
                     \rule{0ex}{15\unitlength}&
                     $-0.005\pm0.009$\\[1ex]
        \hline  
        2 body, total
        \rule{0ex}{15\unitlength}&
                     $-0.021\mp0.009$\\[1ex] 
        \hspace*{\fill}2 body, $g_A^0$
        \rule{0ex}{15\unitlength}&
                     \hspace*{\fill}$-0.021\mp0.009$\\[1ex] 
        \hspace*{\fill}2 body, $g_A^2$
        \rule{0ex}{15\unitlength}&
                     \hspace*{\fill}$-0.0004\pm0.0014$\\[1ex]
        \hline
        \hline
        $a_{\pi d}^\mathrm{exp}$\rule{0ex}{15\unitlength}&
                     $-0.0261\pm0.0005$\\[1ex]
        \hline
      \end{tabular}
      \caption{\textit{One and two body corrections to $a_{\pi^0d}$, choosing
          as input the experimental extraction of $a^+$ from pionic hydrogen
          experiments~\cite{Schroder:1999uq}.}}
      \label{tab:twobody}
    \end{center}
  \vspace*{-4ex}
\end{table}

\absatz A similar analysis can also be made for $D_2+E_2$.  Taking
$a^+_{\mathrm{exp},\pi N}$ from the pionic hydrogen experiment, we find using
the prescription (\ref{eq:D2E2natural})
\begin{equation}
  \label{eq:Zgamma2}
  \calZ=1.0\mp 1.5\;\;
\end{equation}
with a large uncertainty due to the error bar in the value of
$a^+_{\mathrm{exp},\pi N}$. For the central value $\calZ=1.0$ the
contributions from the individual diagrams are $a_{\pi
  d}^{\mathrm{2body}\,(a)}= +0.0005\;\mpi^{-1}$, $a_{\pi
  d}^{\mathrm{2body}\,(b)}=-0.028\;\mpi^{-1}$, $a_{\pi
  d}^{\mathrm{2body}\,(c)}=+0.006\;\mpi^{-1}$.  We note that choosing a
different prescription for removing the $4$ dimensional UV divergences,
$\Gamma\to1$, leads to pion re-scattering contributions numerically deviating
by less than 8\% from those obtained by Beane et al.~\cite{Beane:1998yg}:
$a_{\pi d}^{\mathrm{2body}\,(a)}(\Gamma\to1)= -0.0008\;\mpi^{-1}$, $a_{\pi
  d}^{\mathrm{2body}\,(b)}(\Gamma\to1)=-0.019\;\mpi^{-1}$, $a_{\pi
  d}^{\mathrm{2body}\,(c)}(\Gamma\to1)=-0.002\;\mpi^{-1}$ with
$\calZ(\Gamma\to1)=-0.4\mp1.5$. As the error bars indicate, the magnitude of
$D_2+E_2$ increases as $a^+$ increases. Again, the contribution from the
counter terms is clearly not unnaturally large and is roughly $10-20\%$ of the
dominating pion re-scattering amplitude $a_{\pi d}^{\mathrm{2body}\,(b)}$.
This implies that a $10\%$ uncertainty in the parameters $D_2$ and $E_2$
yields an error of about $1\%$ in the total scattering length. The effect of
the counter terms is thus even smaller than the power counting suggests.
%%%****************************************

One may also assume that the two body counter terms $D_2$ and $E_2$ are
saturated by a mechanism which the phenomenological extraction of $a^+$ from
the pion deuteron scattering length by Ericson et al.~\cite{Ericson:2000md}
can capture correctly. Taking their value $a^+_\mathrm{phen,\,ELT}$, we obtain
$\Lambda^*\approx 280\;\MeV,\;\calZ\approx0.8$. However, such an approach
violates the spirit of self-consistency at the basis of our calculation.

%%%%%%%%%%%%%%% End %%%%%%%%%%%%%%%%%%%
\section{Conclusions}
\setcounter{equation}{0}
\label{sec:conclusions}
%%%%%%%%%%%%%%%%%%%%%%%%%%%%

We presented a calculation of the pion deuteron scattering length in Effective
Field Theory with perturbative pions. In this scheme proposed by Kaplan,
Savage and Wise, knowledge of the iso-scalar pion nucleon scattering length
$a^+$ does not suffice to determine $a_{\pi d}$ directly due to two unknown
short distance effective interactions coupling two nucleons to two pions.
Their strengths $D_2$ and $E_2$ can be subsumed into one physical unknown
which we estimated in the range $\Lambda^*\sim 270\;\MeV$, but with sizeable
error bars.  Employing $\Lambda^*\sim 270\;\MeV$, we can reproduce the
numerical results for the pion re-scattering graphs in the work by Beane et
al.~\cite{Beane:1998yg} who use the Weinberg scheme.  Unfortunately, the
experimental error on $a^+$ is not small enough to constrain $\Lambda^*$ (or
$D_2+E_2$) to a more precise value. The $10\%$ accuracy assigned to our NLO
calculation is not only comparable to the uncertainty induced by experiment in
the sizes of our counter terms. It also masks iso-spin breaking contributions
to $a^+$, which were shown in~\cite{Gasser} not to exceed $7\%$ of the pion
nucleon scattering length. A more accurate measurement of $a^+$ not involving
pion deuteron scattering data, as by the present PSI experiment
R-98.01~\cite{PSIexp}, will substantially reduce the uncertainty.

We can therefore answer the question posed in the Introduction as follows:
Chiral symmetry in combination with already known parameters does not suffice
to determine the pion deuteron scattering length due to two new contact terms
with unknown couplings.  However, once these terms are fixed from the pion
deuteron scattering length they may be used to predict pion deuteron
scattering data at non-zero momentum transfer.

%%%%%%%%%%%%%%%%%%%%%%%%%%%%%%%%%%%%%%%%%%%%%%%%%%%%%%%%%%%%%%%%%%%%%%%%%%%%%%%
%%%%%%%%%%%%%%%%%%%%%%%%%%%%%%%%%%%%%%%%%%%%%%%%%%%%%%%%%%%%%%%%%%%%%%%%%%%%%%%
%%%%%%%%%%%%%%%%%%%%%%%%%%%%%%%%%%%%%%%%%%%%%%%%%%%%%%%%%%%%%%%%%%%%%%%%%%%%%%%

\section*{Acknowledgements}

We are grateful to S.~Beane, T.~Ericson, T.~Hemmert, N.~Kaiser, D.~Phillips,
A.~Thomas, U.~van Kolck and W.~Weise for discussions and useful remarks.
D.~Gotta drew our attention to the ongoing experiments on pionic hydrogen at
PSI. The INT (Seattle) provided generous hospitality during part of this work 
(H.W.G.). We also acknowledge financial support by the DFG Sachbeihilfen GR 
1887/1-1 and GR 1887/2-1 (H.W.G.) and by the Bundesministerium f{\"u}r 
Bildung und Forschung and the Deutsche Forschungsgemeinschaft. 
\newpage

%%%%%%%%%%%%%%%%%%%%%%%%%%%%%%%%%%%%%%%%%%%%%%%%%%%%%%%%%%%%%%%%%%%%%%%%%%%%%%%
%%%%%%%%%%%%%%%%%%%%%%%%%%%%%%%%%%%%%%%%%%%%%%%%%%%%%%%%%%%%%%%%%%%%%%%%%%%%%%%
% Appendix
%
\appendix

%%%%%%%%%%%%%%% App. A: 1 body scattering amplitude %%%%%%%%%%%%%%%%%%%
\section{Calculating the One Body Scattering Amplitude}
\setcounter{equation}{0}
\label{sec:AppA}

As the pion can scatter off the proton or neutron inside the deuteron, the one
body amplitude is (up to relativistic corrections) expected to be given by
twice the physical amplitude $\calA^+$ only, irrespective of the presence of
nested loops at NLO.  We can confirm this formally and turn as an example to
Fig.~\ref{fig:onebody} $(b1)$.  After the integration over the loop energy of
the larger loop, the two loop integral to be performed is using the loop
momentum assignment indicated in Fig.~\ref{fig:onebody} $(b1)$
\begin{equation}
  \int\deintdim{3}{q}\deintdim{4}{l}\frac{1}{(\qv^2+\gamma^2)^2}\;
  \frac{1}{Ml_0-\gamma^2-\frac{\qv^2+(\lv-\qv)^2}{2}}\;
  \frac{\lv^2}{(l_0^2-\lv^2-\mpi^2)^2}\;\;,
\end{equation}
where the second term comes from the nucleon propagator inside the nested
loop, and the last term is the pion propagator. Following the threshold
expansion technique~\cite{BenekeSmirnov,hgpub3}, we identify two poles in the
energy integration of the nested loop: (1) $Ml_0\sim \qv^2,\,\lv^2\sim Q^2$
from the nucleon propagator, and (2) $l_0\sim|\lv|\sim\mpi\sim Q$ from the
pion. In the first case, the pion propagator is expanded because
$l_0\ll|\lv|,\mpi$, so that the pion becomes instantaneous. Since the nucleon
propagates forward in time, this contribution is expected to vanish. The
dimensionally regularised loop integral over $l_0$ is indeed zero as no
external scale is present in the denominators. In the second case, the pion
pole is picked, and the nucleon propagator is expanded into
\begin{equation}
  \frac{1}{Ml_0-\gamma^2-\frac{\qv^2+(\lv-\qv)^2}{2}}\;\to\;
  \frac{1}{Ml_0}\;
  \left[1+\frac{\gamma^2+\frac{\qv^2+(\lv-\qv)^2}{2}}{Ml_0}+\dots\right]\;\;.
\end{equation}
The nucleon propagator becomes static as in \HBCPT, and the two integrals
factorise at LO in the expansion. The (relativistic) corrections are
suppressed by powers of $\frac{Q}{M}\approx\frac{1}{7}$, i.e.~negligible in a
NLO calculation.

We therefore find (\ref{eq:amplitude1body}) as conjectured to NLO, independent
of any deuteron observable:
\begin{equation}
  \calA_{\pi d}^\mathrm{1body}=\frac{4\mpi^2}{\fpi^2}\;
  \left[\frac{\Delta_\mathrm{bare}(\mu)}{2}-\frac{g_A^2}{8M}
         +\frac{g_A^2 (3\mpi-2\mu)}{64 \pi \fpi^2}\right]=2\calA^+\;\;
\end{equation}

%%%%%%%%%%%%%%% App. B: 2 body scattering amplitude %%%%%%%%%%%%%%%%%%%
\section{Renormalising the Two Body Scattering Amplitude}
\setcounter{equation}{0}
\label{sec:AppB}

We will now construct the renormalisation group invariant (and hence physical)
combinations of amplitudes in the two body sector of ${\pi d}$ scattering at
threshold. As mentioned above, the parameter $D_2$ enters in $NN$ scattering
at NLO together with the parameter $C_0^{(0)}$ which respects chiral symmetry.
There, it is also needed to absorb the four dimensional UV divergence stemming
from the one pion exchange diagram.
\begin{figure}[!htb]
  \centerline{\includegraphics*[width=0.9\textwidth]{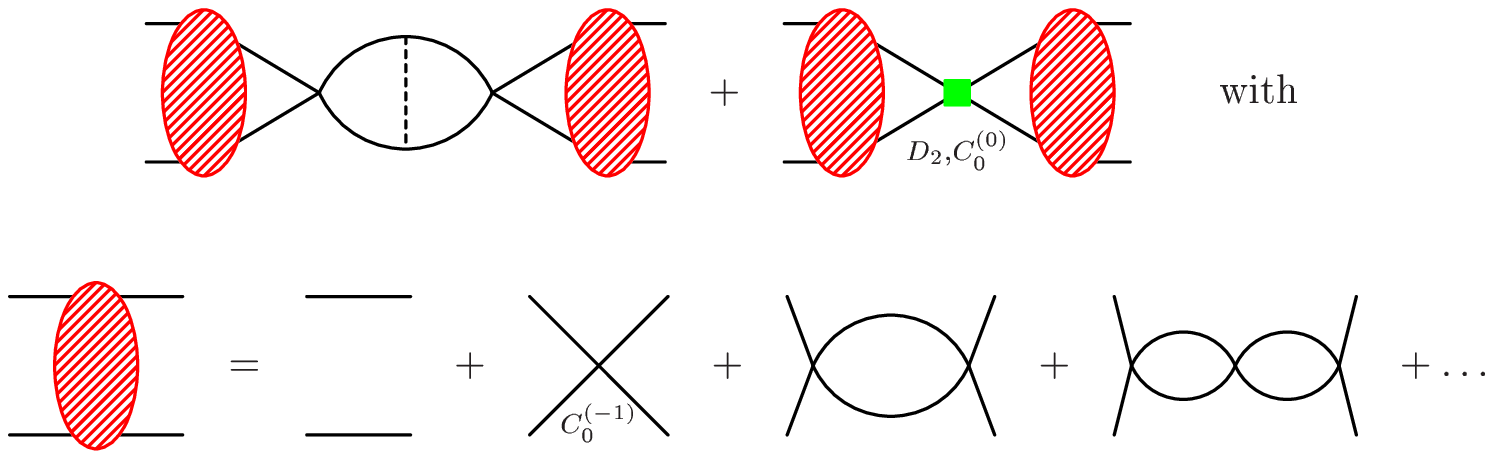}}
    \caption{\textit{The contributions to $NN$ scattering in the
        ${}^3\mathrm{S}_1$ channel at NLO in ENT(KSW) which completely
        determine the renormalisation of $D_2$.}}
    \label{fig:NNscattering}
\end{figure}
We recall~\cite{KSW} that the pertinent terms to renormalise $D_2$ in the
triplet $\mathrm{S}$ wave channel of $NN$ scattering at NLO, depicted in
Fig.~\ref{fig:NNscattering}, are in the centre-of-mass frame between nucleons
with relative momentum $p$ from one pion exchange
\begin{equation}
  \label{eq:NNonepion}
  \calA_{NN,\,\mathrm{bare}}^\mathrm{\pi\;exchange}=
  -\left(\frac{M\calA_{-1}(p)}{4\pi}\right)^2\;
  \frac{g_A^2}{4\fpi^2}\;
  \left[2\left(\mu+\ii p\right)^2-\mpi^2\;
    \left(\Gamma-2\ln\left[\frac{\mpi-2\ii p}{\mu}\right]\right)\right]  
%    \left(C_0^{(-1)}\right)^2\;\frac{g_A^2 M^2}{4(4\pi)^2\fpi^2}
%    \;\left[2\left(\mu+\ii p\right)^2-
%    \mpi^2\;\left(\Gamma-
%                  2\ln\left[\frac{\mpi-2\ii p}{\mu}\right]\right)\right]
\end{equation}
and from the contact interactions involving $D_2$ and $C^{(0)}_{0}$
\begin{equation}
  \label{eq:NNcontact}
  \calA_{NN,\,\mathrm{bare}}^\mathrm{contact}=
  -\left(\mpi^2 D_{2,\;\mathrm{bare}}(\mu)+
    C^{(0)}_{0,\;\mathrm{bare}}(\mu)\right)\;
  \left(\frac{\calA_{-1}(p)}{C_0^{(-1)}}\right)^2\;\;,
%  \mpi^2D_{2,\;\mathrm{bare}}(\mu)\;
%  +\left(C_0^{(-1)}\right)^2\;\frac{M^2}{(4\pi)^2}
%  \left[
%    C^{(0)}_{0,\;\mathrm{bare}}(\mu)\right]
%  (\mu+\ii p)^2\;\;,
\end{equation}
where $C_0^{(-1)}$ is the LO part of the strength of the two nucleon contact
interaction $C_0$, and the connected piece of the LO $NN$ scattering amplitude
(lower line of Fig.~\ref{fig:NNscattering}) is
\begin{equation}
  \label{eq:NNatLO}
  \calA_{-1}(p):=
     -\;\frac{C_0^{(-1)}}{1+\frac{C_0^{(-1)}M}{4\pi}\;(\mu+\ii p)}\;\;.
\end{equation}
The one pion exchange part involving powers of the arbitrary regularisation
parameter $\mu$ comes from the contact piece of one pion exchange which
generates divergences in $3$ dimensions manifest in the PDS scheme. It is
independent of the pion mass, i.e.~unchanged by the chiral limit, and its
regulator dependence can hence be absorbed into the definition of the
renormalised coupling strength
$C^{(0)}_{0,\;\mathrm{R}}:=C^{(0)}_{0,\;\mathrm{bare}}+\frac{g_A^2}{2\fpi^2}$
in accordance with the power counting~\cite{pbhg}. On the other hand, the four
dimensional divergence $\Gamma$ giving rise to the logarithmic dependence on
$\mu$ is chiral symmetry breaking and hence needs to be balanced by the
definition of the bare two nucleon coupling $D_{2,\;\mathrm{bare}}$. $D_2$ is
renormalised by setting
\begin{equation}
  \label{eq:D2ren1}
  D_{2,\;\mathrm{bare}}(\mu) \left(\frac{4\pi}{M C_0^{(-1)}}\right)^2:=
  \frac{g_A^2}{4\fpi^2}
     \left(\Gamma-2\ln\left[\frac{\Lambda^*_{D_2}}{\mu}\right]\right)
     \;\;.
\end{equation}
At LO, $C_0^{(-1)}=-\frac{4\pi}{M}\;(\mu-\gamma)^{-1}$ is determined by
demanding the triplet $\mathrm{S}$ wave to exhibit a pole at the deuteron
binding energy in (\ref{eq:NNatLO}). Therefore, one obtains finally
\begin{equation}
  \label{eq:D2ren}
  D_{2,\;\mathrm{bare}}(\mu) (\mu-\gamma)^2=
  \frac{g_A^2}{4\fpi^2}
     \left(\Gamma-2\ln\left[\frac{\Lambda^*_{D_2}}{\mu}\right]\right)
     \;\;.
\end{equation}
The combination of the divergent one pion exchange diagram and of the contact
interaction depending on the renormalisation group variant parameter
$D_{2,\;\mathrm{bare}}$ is traded for one, renormalisation group invariant
parameter $\Lambda^*_{D_2}$. This dimension-ful, physical number parametrises
the renormalisation group invariant strength of the chiral symmetry breaking
contact interaction between nucleons which does not contain derivatives. It
needs to be determined from experiment or from a microscopic calculation of
$NN$ scattering in QCD. As only a variation of the pion mass can dis-entangle
the effects of $C^{(0)}_{0}$ and $D_2$ in $NN$ scattering at NLO, this process
cannot serve to determine $\Lambda^*_{D_2}$ experimentally.

Inserting the definition (\ref{eq:D2ren}) into the two body amplitude
$\calA_{\pi d}^{\mathrm{2body}}$,
(\ref{eq:barepioncorr}/\ref{eq:bare2bodycorra}) reveals that some dependence
on $\mu$ and on both logarithmic and power law divergences remains: $D_2$
serves only as a partial counter term to the diagrams Fig.~\ref{fig:twobody}
$(a)$ depending on $g_A^2$, and does not affect the divergence of the $g_A^0$
diagram Fig.~\ref{fig:twobody} $(b)$. In the second, analogous step, the
remaining divergences in the two body sector are easily shown to disappear by
setting
\begin{equation}
  \label{eq:E2ren}
    E_{2,\;\mathrm{bare}}(\mu) (\mu-\gamma)^2:=
    \frac{1}{\fpi^2}\;\left(1-\frac{g_A^2}{3}\right)\;
    \left(\Gamma-2\ln\left[\frac{\Lambda^*_{E_2}}{\mu}\right]\right)\;\;,
\end{equation}
where $\Lambda^*_{E_2}$ is another dimension-ful quantity, parameterising the
renormalisation group invariant strength of the quark mass independent
coupling of two nucleons and two pions.

The two body sector of the renormalised amplitude of $\pi d$ scattering at
threshold,
\begin{equation}
  \label{eq:twobodyRGinv1}
  \calA_{\pi d}^\mathrm{2body}=\frac{\mpi^2 \gamma}{\pi\fpi^4}\;
  \left[2\ln\left[\frac{2\gamma}{\Lambda^*_{E_2}}\right]-
    \frac{g_A^2}{6}\left(\half\;\frac{\mpi}{\mpi+2\gamma}+
      \ln\left[\frac{(\mpi+2\gamma)(\Lambda^*_{D_2})^3}{(\Lambda^*_{E_2})^4}
      \right]\right)\right]\;\;,
\end{equation}
contains therefore two un-determined parameters. Pion deuteron scattering at
non-zero momentum transfer will allow to separate the two, and hence also
$C_0^{(0)}$ from $D_2$ in $NN$ scattering, but at present we choose to
represent the two by one common scale
\begin{equation} 
  \label{eq:Lambda}
  (\Lambda^*)^{\frac{g_A^2}{12}-1}:=(\Lambda^*_{D_2})^{-\frac{g_A^2}{4}}
                                \;(\Lambda^*_{E_2})^{\frac{g_A^2}{3}-1}\;\;.
\end{equation}
The final answer for the two body scattering amplitude in terms of
renormalised quantities is therefore given by (\ref{eq:twobodyRGinv})
\begin{equation}
  \calA_{\pi d}^\mathrm{2body}=\frac{\mpi^2 \gamma}{\pi\fpi^4}\;
  \left[2\ln\left[\frac{2\gamma}{\Lambda^*}\right]-
    \frac{g_A^2}{6}\left(\half\;\frac{\mpi}{\mpi+2\gamma}+
      \ln\left[\frac{(\mpi+2\gamma)}{\Lambda^*}\right]
    \right)\right]\;\;.
\end{equation}
The well known fact that dimensional regularisation (together with the
employed renormalisation scheme) preserves chiral symmetry manifests itself in
the observation that all appearing divergences are renormalised manifestly
chirally invariant.
%%%%%%%%%*****************************************

\newpage

%%%%%%%%%%%%%%%%%%%%%%%%%%%%%%%%%%%%%%%%%%%%%%%%%%%%%%%%%%%%%%%%%%%%%%%%%%%%%%%
%%%%%%%%%%%%%%%%%%%%%%%%%%%%%%%%%%%%%%%%%%%%%%%%%%%%%%%%%%%%%%%%%%%%%%%%%%%%%%%
%%%%%%%%%%%%%%%%%%%%%%%%%%%%%%%%%%%%%%%%%%%%%%%%%%%%%%%%%%%%%%%%%%%%%%%%%%%%%%%

%\end{fmffile}
\end{document}